# Ferroelectrovalley in Two-Dimensional Multiferroic Lattices

Jiangyu Zhao[1], Yangyang Feng[1], Ying Dai*[,1], Baibiao Huang[1], Yandong Ma*[,1]

[1]School of Physics, State Key Laboratory of Crystal Materials, Shandong University, Shandanan Street 27, Jinan 250100, China

*Corresponding author: daiy60@sdu.edu.cn (Y.D.); yandong.ma@sdu.edu.cn (Y.M.)

**Abstract**

Engineering valley index is essential and highly sought for valley physics, but currently it is exclusively based on the paradigm of the challenging ferrovalley with spin-orientation reversal under magnetic field. Here, an alternative strategy, i.e., the so-called ferroelectrovalley, is proposed to tackle the insurmountable spin-orientation reversal, which reveres valley index with the feasible ferroelectricity. Using symmetry arguments and tight-binding model, the $C_2$ rotation is unveiled to be able to take the place of time reversal for operating valley index in two-dimensional multiferroic kagome lattices, which enables the ferroelectricity-engineered valley index, thereby generating the concept of ferroelectrovalley. Based on first-principles calculations, this concept is further demonstrated in the breathing kagome lattice of single-layer $Ti_3Br_8$, wherein ferroelectricity couples the breathing process. These findings open a new direction for valleytronics and two-dimensional materials research.

**Introduction**

In many crystalline solids, it often happens that the conduction or valence bands have two or more local energy extrema in the momentum space. This leads to an additional valley degree of freedom for low-energy carriers [1-4], which is robust against low-energy phonons and smooth deformation due to the large separation in momentum space [5-7]. Similar to spin for spintronics, the possibility of using valley degree to store and encode information leads to the conceptual electronic applications known as valleytronics [1,2]. In recent years, with the discovery of valley degree in two-dimensional (2D) lattice [5,8-12], the valley physics has attracted broad interest from the perspectives of both fundamental research and device applications, and has been investigated intensively [13-19].

In view of realizing valleytronic devices, it is necessary to achieve effective control of valley index [20,21]. Currently, the valleytronics research has been exclusively established in the paradigm of time-reversal ($T$) symmetry connected ferrovalley [20,22], wherein the valley index is manipulated through spin-orientation reversal under inversing magnetic field. However, such spin-orientation reversal in 2D lattice is extremely difficult, which severely limits the ferrovalley paradigm for valleytronics. Different from magnetic field, the static control by gate electric field is most desirable among all static means, because of its advantages in compactness and compatibility with the existing semiconductor technology [23]. Through highly desirable, under the existing ferrovalley paradigm, the gate field cannot couple with the valley index [10,24]. These rise an outstanding challenge for valleytronics research.

In this work, going beyond the well-established ferrovalley paradigm, we propose an alternative strategy, i.e., the so-called ferroelectrovalley, to circumvent this challenge, wherein the valley index switch is related to ferroelectric physics. Our symmetry and tight-binding model analysis unveil that the $C_2$ rotation is capable of taking over $T$ reversal to reverse valley index in two-dimensional multiferroic kagome lattices. This suggests the intriguing feasibility of coupling valley index with ferroelectricity, thus generating the concept of ferroelectrovalley. Using first-principles calculations, we further demonstrate this concept in the breathing kagome lattice of single-layer (SL) Ti3Br8, wherein the ferroelectricity associated breathing process connects $C_2$ rotation. Our work not only greatly enriches the valleytronics and 2D materials research, but also uncover a unique and general mechanism to control valley index by electric field.

**Results and Discussion**

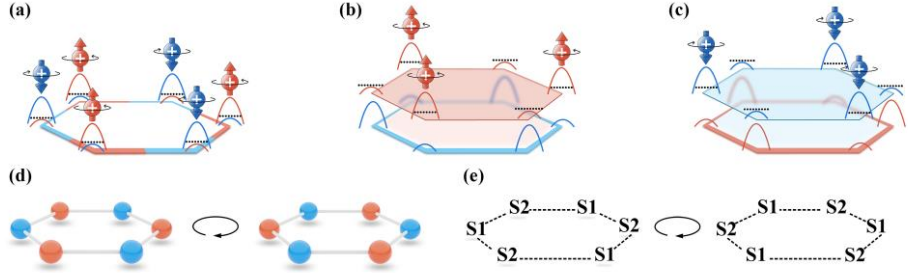

**Figure 1**. Schematic diagrams of the low-energy band dispersions of ferrovalley system (a) without $H_{ex}$ and (b, c) with $H_{ex}$. (c) is same as (b), but with opposite spin-orientation. Black dotted lines indicate the Fermi level. Schematic diagrams of the evaluations of (d) lattice and (e) states in the Brillouin zone under $C_2$ rotation.

The proposed mechanism is rooted in 2D hexagonal systems with *T*-symmetry breaking and valley physics. The Hamiltonian for such systems near the ±K valleys can be expressed as: $H = H_0 + H_{ex} + H_{SOC}$. The second term is the inherent exchange interaction of magnetic ions, which can be written as:

$$H_{ex} = -\boldsymbol{S} \cdot \boldsymbol{m} \qquad (1)$$

Here, $\boldsymbol{S}$ is spin angular momentum and $\boldsymbol{m}$ is effective exchange splitting. The third term is the spin-orbit coupling (SOC) effect, which is essential for valley physics and can be written as:

$$H_{SOC}(\boldsymbol{k}) = \frac{\lambda}{2}\boldsymbol{S} \cdot \boldsymbol{L} \qquad (2)$$

Here, $\lambda$ is the SOC parameter, $\boldsymbol{L}$ is orbital angular momentum, and $\boldsymbol{k}$ is the reciprocal lattice vector. Note $\boldsymbol{L}(-\boldsymbol{k}) = -\boldsymbol{L}(\boldsymbol{k})$, for a given spin channel, the energy shifts caused by SOC around valleys follow the relation:

$$H_{SOC}^{\uparrow/\downarrow}(-\boldsymbol{k}) = -H_{SOC}^{\uparrow/\downarrow}(\boldsymbol{k}) \qquad (3)$$

where ↑/↓ denote spin up/down. While for different spin channels, we can establish the following relation:

$$H_{SOC}^{\downarrow}(k) = \frac{\lambda}{2}(S^{\downarrow}) \cdot L = \frac{\lambda}{2}(S^{\uparrow}) \cdot (-L) = H_{SOC}^{\uparrow}(-k) \quad (4)$$

According to equations (3, 4), when excluding $H_{ex}$, the +K and -K valleys are degenerate in energy, and they are from opposite spin channels [**Figure 1(a)**]. By including $H_{ex}$, as illustrated in **Figure 1(b)**, the degeneracy of the valleys is lifted, resulting in valley polarization. Then, the carriers can be populated at one given valley. Under the paradigm of ferrovalley [25-27], through inverting the spin orientation by applying magnetic field, the valley polarization is reversed [**Figure 1(c)**], thereby realizing the switch of valley index of the carriers. Going beyond the spin-orientation reversal of the ferrovalley paradigm, intriguingly, we find that inversion ($I$) operation can also lead to the switch of valley index of the carriers. Based on equations (1, 2), under the $I$ operation, we obtain $H_{ex} = I[H_{ex}]$, and $H_{soc}(-k) = I[H_{SOC}(k)]$. This along with $H_0(-k) = H_0(k)$ yields the relation of $I[H(k)] = H(-k)$. It indicates that with performing $I$ operation on the lattice, the states at $-k$ and $k$ points would be reversed [**Figure 1(d)** and **1(e)**]. In single atomic lattice, the $I$ operation is equivalent to the $C_2$ rotation. Therefore, analogous to switching spin orientation, the $C_2$ rotation can also guarantee the valley index reversal.

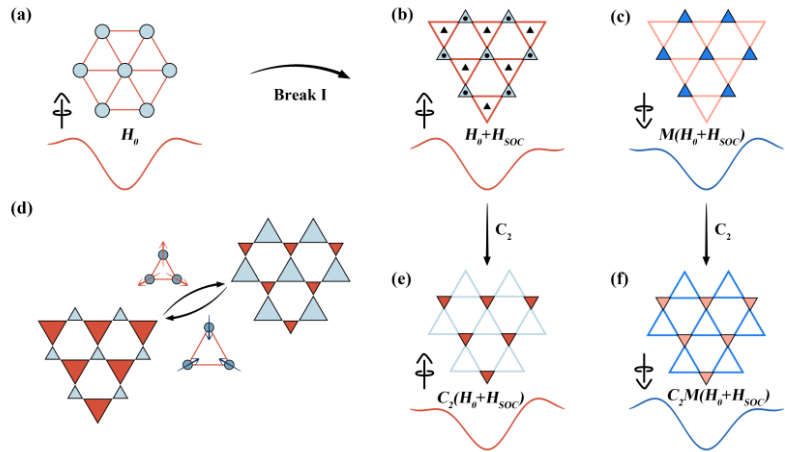

**Figure 2**. (a) 2D triangular lattice and its corresponding low-energy band dispersion. (b) 2D kagome lattice evaluated from (a) and its corresponding low-energy band dispersion. Black dots and triangles represent the $\Theta_I$ and $\Theta_{II}$ $C_3$ axes respectively. (c)

is the same as (b) but with opposite spin orientation. (e) is the same as (b) but under a $C_2$ rotation. (f) is the same as (e) but with opposite spin orientation. In (b, c, e, f), black arrows represent the direction of spin orientation. (d) Schematic diagram of equivalent $C_2$ rotation.

As operating $C_2$ rotation is impractical, we alternatively resort to equivalent $C_2$ rotation, i.e., the diffusionless phase transformation. In the following, we start from the triangular lattice formed by magnetic atoms to investigate the diffusionless phase transformation [**Figure 2(a)**]. A simple model for this lattice can be written as:

$$H' = -\sum_{\langle i,j \rangle} t_{ij} a_i^\dagger a_j + h.c. + \epsilon + H_{SOC} \tag{5}$$

Here, $t_{ij}$ is the hopping term, $\epsilon$ is the energy of orbital and inherent exchange interaction energy of magnetic ions, and $\langle i,j \rangle$ indicates sum over nearest neighbors. For more detail, please see the Supplement Information. The low-energy band dispersion obtained from equation (5) is shown in the low panel of **Figure 2(a)**. Due to the protection of *I*-symmetry, the valley physics is absent. To break the *I*-symmetry, clusters with $C_3$ symmetry is introduced to replace atoms in the triangular lattice, as shown in **Figure 2(b, c)**, which generates a kagome lattice with valley polarization.

Due to the $C_3$ symmetry, each cluster contains $n = 3m$ atoms, which thus can be divided into three equivalent groups. Each group has $m$ atoms, and the three groups are connected by $C_3$ rotation. Intriguingly, there are two types of $C_3$ axes for the kagome lattice [**Figure 2(b)**], i.e., $\Theta_\text{I}$ and $\Theta_\text{II}$. In this sense, as illustrated in **Figure 2(d)**, the diffusionless phase transformation can be realized by expelling the three groups of atoms from $\Theta_\text{I}$ axis and propelling them to their corresponding nearest-neighboring $\Theta_\text{II}$ axis. As a result, three groups from different loosened clusters reform a new cluster around the $\Theta_\text{II}$ axis, corresponding to a breathing process. The newly resultant kagome lattice can be considered as operating a $C_2$ rotation on the old one. According to the low-energy effective model of equation (S8), this equivalent $C_2$ rotation indeed can reverse the valley index of the populated carriers, as shown in **Figure 2(e, f)**. It is worthy emphasizing that with considering the coordinated atoms, such breathing process in the kagome lattice can be closely correlated to ferroelectric phase transition [28,29]. This suggests that the valley index of the populated carriers can be controlled by electric field, thereby generating the concept of ferroelectrovalley.

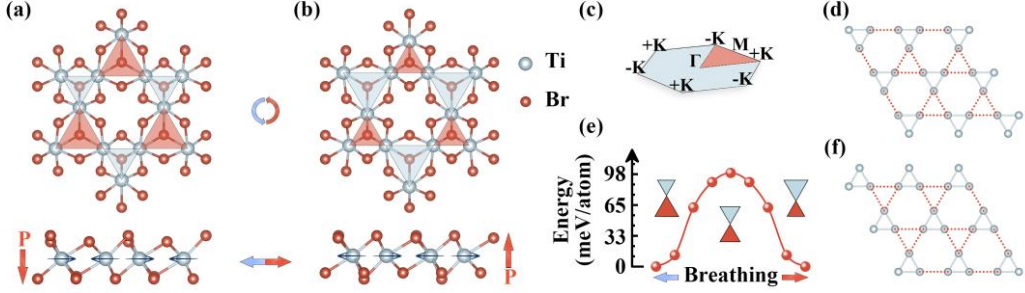

**Figure 3.** (a, b) Crystal structures of the two FE states of SL $Ti_3Br_8$ from the top and side views. In (a, b), dark arrows indicate the directional displacement of magnetic atoms during the breathing process, and the orange arrows with the letter P marks the electric polarization directions. (c) Schematic diagram of Brillouin zone with marking the high-symmetry points. (d, f) Schematic diagram of clusters for the two FE states of SL $Ti_3Br_8$. (e) Energy profiles for FE switching of SL $Ti_3Br_8$. In (e), the size variation of the red and blue triangles illustrates the breathing process.

One candidate system for realizing this mechanism is SL $Ti_3Br_8$. **Figure 3(a)** shows the crystal structure of $Ti_3Br_8$, which possesses a hexagonal kagome lattice with P3m1 (No. 156) space group. The optimized lattice parameter is 7.23 Å, which agrees well with the previous study [28]. Each unit cell consists of eight Br and three Ti atoms, with Ti atomic layer sandwiched between two Br atomic layers. Each Ti atom is coordinated by six Br atoms in a distorted octahedral geometry. In the Ti atomic lattice, every three Ti atoms form a $Ti_3$ trimer, as represented by small blue triangles in **Figure 3(a)**. This forms a kagome lattice, and results in two kinds of Ti-Ti bonds. In the $Ti_3$ trimer, the Ti-Ti bond length is 3.12 Å, while for the Ti-Ti bond in the large red triangle, it is 4.10 Å.

By focusing on the structural symmetry of SL $Ti_3Br_8$, we can see that both the *I* and horizontal mirror symmetries are absent. Quietly naturally, this gives rise to an electric polarization of 0.7 pC/m along the z-direction. Such symmetry broken is closely related to the distortion of the octahedral geometry of $TiBr_6$. In fact, the distortion can also occur in the opposite direction, resulting in another phase of SL $Ti_3Br_8$, as shown in **Figure 3(b)**. Obviously, these two phases are degenerate in energy and can be considered as two ferroelectric (FE) states of SL $Ti_3Br_8$, i.e., phases I and II, respectively.

As shown in **Figure 3(a, b)**, the FE transition in SL $Ti_3Br_8$ is accompanied with a breathing process. In detail, under FE transition, the three Ti atoms from one $Ti_3$ trimer expel from the center of triangles, and the three atoms from three different neighboring $Ti_3$ trimers clusters to a new $Ti_3$ trimer around the center of another triangle. This can be simply considered as scaling up the blue triangles and scaling down the red triangles, or vice versa, which corresponds to a breathing process. The corresponding arrangements of Ti atoms under phase I and phase II are depicted in **Figure 3(d, f)**, from which we can see the ferroelectric phase transition can be regarded as an equivalent $C_2$ rotation. This is in good agreement with the above model analysis shown in **Figure 2(d)**.

To get more insight into the ferroelectricity in SL $Ti_3Br_8$, we investigate the FE switching process using the climbing image nudged elastic band method (CL-NEB) method. As shown in **Figure 3(e)**, phases I and II can be switched to each other with an energy barrier of 99.3 meV/atom. This is much lower than $FeO_2H$ [30], SiN [31] and $CrNCl_2$ [32], and comparable with that of $Au_2Cl_2Te_4$ [33], which suggests the feasibility of the ferroelectricity in SL $Ti_3Br_8$.

In addition to ferroelectricity, SL $Ti_3Br_8$ also harbors spin polarization. The valence electric configuration of Ti atom is $3d^24s^2$. Before clustering $Ti_3$ trimer, because of the octahedral crystal field, the $d$ orbitals of Ti atoms split into $e_g$ and $t_{2g}$. After forming $Ti_3$ trimer, the $t_{2g}$ orbitals split into $1e$, $1a_1$, $2e$, $2a_1$, $3e$, and $1a_2$, see **Figure S1**. e, $a_1$, $a_2$ are irreducible representation in $C_{3v}$ point group, among which e is the basis function of $(d_{x2-y2}, d_{xy})$ $(d_{xz}, d_{yz})$, and $a_1$ is the basis function of $d_{z2}$. Ti atoms in each $Ti_3$ trimer share 4 electrons, yielding a magnetic moment of 2 $\mu_B$. As a result, SL $Ti_3Br_8$ is spin polarized.

Although it prefers in-plane orientation, its magnetic anisotropy energy is only 0.14 meV/unit cell, which is easy to be manipulated by magnetic field. To estimate the magnetic ground state of SL $Ti_3Br_8$, we consider a ferromagnetic (FM) state and three antiferromagnetic (AFM) states (AFM1, AFM2, and AFM3). The considered magnetic configurations are shown in **Figure S2**. Our calculation shows that the FM state is the ground state. In detail, the energy of FM state is 2.827, 2.830, and 2.830 eV/unit cell lower than AFM1, AFM2, and AFM3 states respectively.

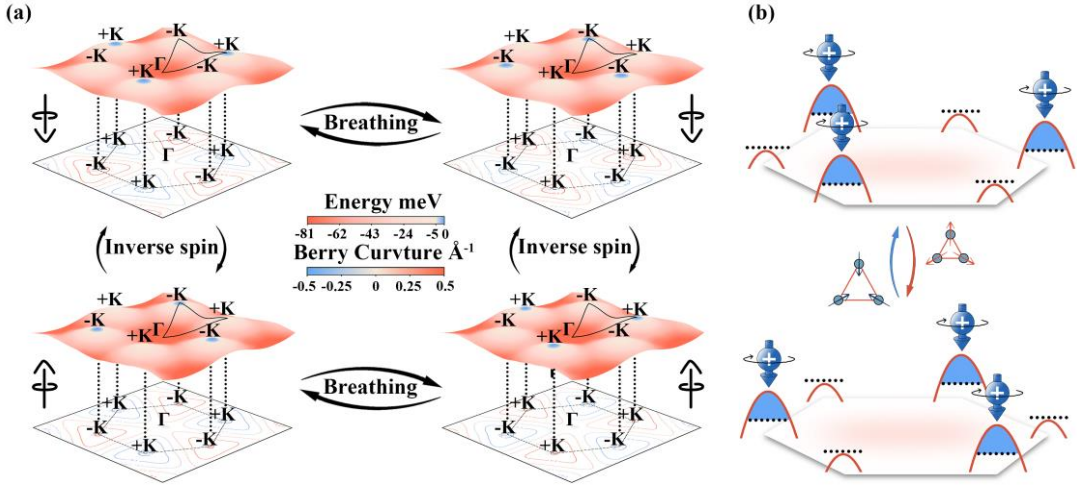

**Figure 4**. (a) Dispersion of the top valence band and the Berry curvature of SL $Ti_3Br_8$. The 3D curved surface represents the band structures, while the contour map illustrates the Berry curvature. In (a), the left/right two subfigures represent phase I/II; the top/bottom two subfigures correspond to up/down spin-orientation. The color of the surface and contour map represents the value of energy and Berry curvature, respectively. The Fermi energy is shifted to zero. The black arrows represent the direction of spin orientation. (b) Schematic diagram of ferroelectrovalley. In (b), the black dash line denotes the Fermi energy and the region filled with blue color implies the populated carriers; the red/blue balls with arrows represent the up/down-spin holes.

The band dispersions of top valence band and the corresponding Berry curvatures of SL $Ti_3Br_8$ are shown in **Figure 4(a)**. The top valence band is an isolated band and separated by a band gap of 206 meV from the deeper bands. It is contributed by the same spin channel and exhibits two inequivalent valleys located at the ±K points. And from the contour map of **Figure 4(a),** we can see that, due to the simultaneous broken of *I*- and *T*-symmetries in SL Ti$_3$Br$_8$, the Berry curvatures of the two inequivalent valleys exhibit opposite signs.

For phase I, as shown in upper-left panel of **Figure 4(a)**, the +K valley lies higher in energy than the -K valley, giving rise to a valley polarization of 5 meV. When imposing a FE transition (i.e., phase II), the kagome lattice breathes, leads to that the +K valley shifts below the -K valley; see upper-right panel of **Figure 4(a)**. In other word, the valley polarization is reversed through FE switching. This indicates a strong coupling between the valley polarization and ferroelectricity. This scenario is also applicable for the cases with opposite spin orientation; see lower panels of **Figure 4(a)**.

In this sense, the concept of ferroelectrovalley can be achieved in SL Ti$_3$Br$_8$. We here take the cases with up spin-orientation as examples to illustrate the ferroelectrovalley. As schematically displayed in **Figure 4(b)**, when shifting the Fermi level between the two valleys, the carriers are populated at +K valley. That is to say, the valley index of the carriers is $\tau = 1$. With applying electric field to induce the FE transition, the carriers are populated at the -K valley, indicating the valley index of the carriers is switched to $\tau = -1$. Therefore, different from the ferrovalley paradigm, the valley index of carriers can be controlled by electric field, demonstrating the concept of ferroelectrovalley.

**Conclusion**

To summarize, going beyond the existing ferrovalley paradigm relayed on the annoying spin-orientation reversal, we propose an alternative strategy, i.e., the concept of ferroelectrovalley, to manipulate the valley index. Based on symmetry arguments and tight-binding model, we unveil that the $C_2$ rotation is able to take the place of *T*-reversal for operating valley index in two-dimensional multiferroic kagome lattices. This guarantees the ferroelectricity-engineered valley index, i.e., the concept of ferroelectrovalley. Our first-principles calculations further demonstrate this concept in the breathing kagome lattice of single-layer Ti$_3$Br$_8$, wherein ferroelectricity couples the breathing process.

**Computational Methods**

Our first-principles calculations are performed based on the density-functional theory (DFT) [34] as implemented in the Vienna ab initio simulation package (VASP) [35]. We employ the generalized gradient approximation in the form of Perdew-Burke-Ernzerhof functional [36] to describe the exchange-correlation potential. A 7×7×1 grid is employed to sample the Brillouin zone. The cutoff energy is set to 600 eV and the convergence criterion of total energy is set to $10^{-6}$ eV. We fully relax both the position of all atoms and the lattice parameter until the force on each atom is less than $10^{-2}$ eV/Å. We employ CL-NEB method [37] to study the FE transition. We use the VASPBERRY code [38] to calculate the Berry curvature.

**Acknowledgment**

This work is supported by the National Natural Science Foundation of China (Nos. 12274261 and 12074217), Shandong Provincial QingChuang Technology Support Plan (No. 2021KJ002), and Taishan Young Scholar Program of Shandong Province.

**Supplement Information**

See Supplement Material for details of tight-binding model, *d*-orbital splitting of $Ti_3$ trimers, and magnetic configurations.

**Reference**

[1] J. R. Schaibley, H. Yu, G. Clark, P. Rivera, J. S. Ross, K. L. Seyler, W. Yao, and X. Xu, Nat. Rev. Mater. 1, 16055 (2016).

[2] T. Zhou, J. Zhang, H. Jiang, I. Žutić, and Z. Yang, npj Quantum Mater. 3, 39 (2018).

[3] T. Zhou, S. Cheng, M. Schleenvoigt, P. Schüffelgen, H. Jiang, Z. Yang, and I. Žutić, Phys. Rev. Lett. 127, 116402 (2021).

[4] D. Xiao, W. Yao, and Q. Niu, Phys. Rev. Lett. 99, 236809 (2007).

[5] H. Lu, W. Yao, D. Xiao, and S. Shen, Phys. Rev. Lett. 110, 016806 (2013).

[6] D. Xiao, G. Liu, W. Feng, X. Xu, and W. Yao, Phys. Rev. Lett. 108, 196802 (2012).

[7] R. Peng, Z. He, Q. Wu, Y. Dai, B. Huang, and Y. Ma, Phys. Rev. B 104, 174411 (2021).

[8] S. Zhang, D. Shao, Z. Wang, J. Yang, W. Yang, and E. Y. Tsymbal, Phys. Rev. Lett. 131, 246301 (2023).

[9] L. Ju, Z. Shi, N. Nair, Y. Lv, C. Jin, J. Velasco, C. Ojeda-Aristizabal, H. A. Bechtel, M. C. Martin, A. Zettl, J. Analytis, and F. Wang, Nature 520, 650 (2015).

[10] X. Ma, L. Sun, J. Liu, X. Feng, W. Li, J. Hu, and M. Zhao, Phys. Status Solidi RRL 14, 2000008 (2020).

[11] Z. He, R. Peng, X. Feng, X. Xu, Y. Dai, B. Huang, and Y. Ma, Phys. Rev. B 104, 075105 (2021).

[12] S. A. Vitale, D. Nezich, J. O. Varghese, P. Kim, N. Gedik, P. Jarillo-Herrero, D. Xiao, and M. Rothschild, Small 14, 1801483 (2018).

[13] K. F. Mak, K. He, J. Shan, and T. F. Heinz, Nat. Nanotechnol. 7, 494 (2012).

[14] B. A. Foutty, J. Yu, T. Devakul, C. R. Kometter, Y. Zhang, K. Watanabe, T. Taniguchi, L. Fu, and B. E. Feldman, Nat. Mater. 22, 731 (2023).

[15] G. Yu, J. Ji, C. Xu, and H. J. Xiang, Phys. Rev. B 109, 075434 (2024).

[16] M. A. Mueed, M. S. Hossain, I. Jo, L. N. Pfeiffer, K. W. West, K. W. Baldwin, and M. Shayegan, Phys. Rev. Lett. 121, 036802 (2018).

[17] W. Zhou, G. Zheng, Z. Wan, T. Sun, A. Li, and F. Ouyang, Appl. Phys. Lett. 123, 143101 (2023).

[18] Y. Zhang, K. Shinokita, K. Watanabe, T. Taniguchi, Y. Miyauchi, and K. Matsuda, Adv. Funct. Mater. 31, 2006064 (2021).

[19] D. Degli Esposti, L. E. A. Stehouwer, Ö. Gül, N. Samkharadze, C. Déprez, M. Meyer, I. N. Meijer, L. Tryputen, S. Karwal, M. Botifoll, J. Arbiol, S. V. Amitonov, L. M. K. Vandersypen, A. Sammak, M. Veldhorst, and G. Scappucci, npj Quantum Inf. 10, 32 (2024).

[20] D. MacNeill, C. Heikes, K. F. Mak, Z. Anderson, A. Kormányos, V. Zólyomi, J.

Park, and D. C. Ralph, Phys. Rev. Lett. 114, 037401 (2015).

[21] C. W. J. Beenakker, N. V. Gnezdilov, E. Dresselhaus, V. P. Ostroukh, Y. Herasymenko, İ. Adagideli, and J. Tworzydło, Phys. Rev. B 97, 241403 (2018).

[22] J. J. Lu, R. Liu, F. F. Yue, X. W. Zhao, G. C. Hu, X. B. Yuan, and J. F. Ren, J. Phys. Chem. Lett. 14, 132 (2023).

[23] E. Y. Tsymbal, Nat. Mater. 11, 12 (2012).

[24] D. Zhang, A. Li, X. Chen, W. Zhou, and F. Ouyang, Phys. Rev. B 105, 085408 (2022).

[25] W. Tong, S. Gong, X. Wan, and C. Duan, Nat. Commun. 7, 13612 (2016).

[26] J. Zhao, Y. Qi, C. Yao, and H. Zeng, Appl. Phys. Lett. 124, 093103 (2024).

[27] H. Cheng, J. Zhou, W. Ji, Y. Zhang, and Y. Feng, Phys. Rev. B 103, 125121 (2021).

[28] Y. Li, C. Liu, G. Zhao, T. Hu, and W. Ren, Phys. Rev. B 104, L060405 (2021).

[29] D. Hu, H. Ye, N. Ding, K. Xu, S. Wang, S. Dong, and X. Yao, Phys. Rev. B 109, 014433 (2024).

[30] S. Shen, X. Xu, B. Huang, L. Kou, Y. Dai, and Y. Ma, Phys. Rev. B 103, 144101 (2021).

[31] Y. Feng, T. Zhang, Y. Dai, B. Huang, and Y. Ma, Appl. Phys. Lett. 120, 193102 (2022).

[32] W. Shen, Y. Pan, S. Shen, H. Li, S. Nie, and J. Mei, Chin. Phys. B. 30, 117503 (2021).

[33] M. Kruse, U. Petralanda, M. N. Gjerding, K. W. Jacobsen, K. S. Thygesen, and T. Olsen, npj Comput. Mater. 9, 45 (2023).

[34] W. Kohn, and L. J. Sham, Phys. Rev. 140, A1133 (1965).

[35] G. Kresse, and J. Furthmüller, Phys. Rev. B 54, 11169 (1996).

[36] J. P. Perdew, K. Burke, and M. Ernzerhof, Phys. Rev. Lett. 77, 3865 (1996).

[37] G. Henkelman, B. P. Uberuaga, and H. Jónsson, J. Chem. Phys. 113, 9901 (2000).

[38] S. Kim, H. Kim, S. Cheon, and T. Kim, Phys. Rev. Lett. 128, 046401 (2022).